# Size-controlled quantum dots reveal the impact of intraband transitions on high-order harmonic generation in solids

Kotaro Nakagawa[1], Hideki Hirori[1,*], Shunsuke A. Sato[2,3], Hirokazu Tahara[1], Fumiya Sekiguchi[1], Go Yumoto[1], Masaki Saruyama[1], Ryota Sato[1], Toshiharu Teranishi[1], and Yoshihiko Kanemitsu[1,*]

[1]*Institute for Chemical Research, Kyoto University, Uji, Kyoto 611-0011, Japan*

[2]*Center for Computational Sciences, University of Tsukuba, Tsukuba 305-8577, Japan*

[3]*Max Planck Institute for the Structure and Dynamics of Matter, Luruper Chaussee 149, 22761 Hamburg, Germany*

*Corresponding author: hirori@scl.kyoto-u.ac.jp
*Corresponding author: kanemitu@scl.kyoto-u.ac.jp

**Since the discovery of high-order harmonic generation (HHG) in solids[1–3], much effort has been devoted to understanding its generation mechanism and both interband and intraband transitions are known to be essential[1–10]. However, intraband transitions are affected by the electronic structure of a solid, and how they contribute to nonlinear carrier generation and HHG remains an open question. Here, we use mid-infrared laser pulses to study HHG in CdSe and CdS quantum dots (QDs), where quantum confinement can be used to control the intraband transitions. We find that both the HHG intensity per excited volume and the generated carrier density increase when the average QD size is increased from about 2 nm to 3 nm. We show that the reduction of the subband gap energy in larger QDs enhances intraband transitions, and this in turn increases the rate of photocarrier injection by coupling with interband transitions, resulting in enhanced HHG.**

The transitions between the valence band (VB) and the conduction band (CB) are responsible for photon absorption in a semiconductor[11]. However, a strong nonlinear optical response is not necessarily governed only by such interband transitions, and thus we need to understand the interplay between interband and intraband transitions. In the case that the electric field $E(t)$ of a laser accelerates an electron in the band of a solid, which constitutes an intraband transition, the temporal change in the electron wavenumber $k(t)$ can be described by $\hbar dk(t)/dt = eE(t)$, where $e$ is the electron charge and $\hbar$ is the reduced Planck constant. Depending on the degree of the change in $k$, two regimes can be considered: conventional and extreme nonlinear optical phenomena. In the conventional regime (that is, under resonant or near-resonant conditions), the optical field induces only small changes in the wavenumber, leading to a dominant contribution of the interband transitions[12–14]. In the extreme nonlinear regime where the excitation photon energy $\hbar\omega_0$ is much smaller than the bandgap energy $E_g$, efficient carrier acceleration is possible without damaging the sample by excessive carrier generation. Thus, it is possible to use strong long-wavelength laser fields that induce large changes in the wavenumber. In this case, intraband transitions play a major role in the extreme nonlinear dynamics of the system.

Recently, a rather unexpected role of intraband transitions was found; carrier injection into the CB of GaAs is enhanced by coupling of interband and intraband transitions[15]. Also, it has been shown that nonlinear-carrier-generation processes in the extreme nonlinear regime play an important role in the modification of optical and electric properties of solids in the ultrafast time scales[16–19]. These phenomena are related to high-order harmonic generation (HHG) in solids because, in addition to interband transition, nonlinear intraband transition is considered to be responsible for the elementary excitation process behind this phenomenon. However, the relation between nonlinear carrier generation and HHG has not yet been experimentally studied, and thus it has remained elusive how intraband transitions correlate with the electronic structure of a solid and eventually with nonlinear carrier generation and HHG. To clarify the impact of intraband transitions on the extreme nonlinear optical phenomena, it is necessary to study it from materials in which the electronic band structure and thus intraband transition can be freely controlled. Because the nature of the electronic bands of quantum dots (QDs) can be continuously tuned from atom-like discrete states to a solid-state band continuum simply by changing their size, without changing the constituting elements[20–23], they are ideal materials to examine the role of intraband transitions in carrier generation and its relation to HHG.

Here, we studied HHG in CdSe and CdS QD films. The dashed lines on the left-hand side of Fig. 1a schematically illustrate the energy levels of a QD; $E_g$ is the QD bandgap energy and $\Delta_{sub}$ is the first subband gap energy. Both parameters depend on the QD diameter. The middle panel of Fig. 1a shows the absorption spectra of CdSe QDs with different diameters (Supplementary Information I). Owing to the strong quantum confinement in these small QDs, $E_g$ is larger than in bulk by up to several hundred milli-electronvolts (the bandgap energies of bulk CdSe and CdS are 1.75 eV and 2.58 eV, respectively), and the electronic states become discrete[24–27]. The right-hand side of Fig. 1a shows the excitation and detection geometry used to measure the HHG emission spectra of CdSe QD film (see the Methods for details) and the typical transmission electron microscope (TEM) images. Figure 1b shows the high-order harmonic (HH) intensities per excited volume as a function of the photon energy $\hbar\omega$, $I_{HHG}(\hbar\omega)$, for CdSe and CdS QD films under excitation with linearly polarized mid-infrared (MIR) light. The spectra extend from the visible to the ultraviolet region and the peaks correspond to the 7th–13th orders. We find that $I_{HHG}(\hbar\omega)$ tends to increase as the average QD diameter $d$ increases, and it increases abruptly in the range from 2 to 3 nm.

Figure 2 provides data on CdSe (red circles) and CdS (blue squares): it shows the integrated HH peak intensity of each order ($I_h$, where $h$ is the harmonic order) as a function of $d$ to clarify the QD size dependence of the HH intensity. For CdSe QDs, $I_h$ substantially increases with $d$ in the range $d \approx 1.8$–3.8 nm. For example, $I_7$ increases by a factor of about 100 from $d = 2.1$ to 3.8 nm. Note that the $E_g$ of CdSe QDs with $d$ in the range 2.1–3.8 nm changes from 2.6 eV to 2.1 eV, but the absorption spectra in Fig. 1a indicate that the required number of photons in the multiphoton absorption process for the VB–CB transition remains almost the same (i.e., the band edge lies in the region near $h = 7$ or 2.48 eV). Although the bandgap energy for $d = 2.4$ nm lies closer to the 7th resonant multiphoton absorption than that for 2.8 nm, the HH intensities of the smaller QDs are much smaller. Also, an almost constant behavior in the range $d \approx 3.8$–14 nm (the inset of Fig. 2a) is observed despite of the difference in the order of multiphoton absorption. In addition, the excitation intensity dependence shows that the HHG mechanism in CdSe QDs under these excitation conditions is non-perturbative (Supplementary Information II and III). These results show that the size dependence of $I_h$ cannot be simply explained by the difference in the multiphoton absorption process considering only interband transitions.

To obtain the relation between the actual carrier density generated by the MIR pump pulse and HHG, we measured the transient absorption (TA) change of the CdSe QD films in experiments with an MIR pump and a white-light probe. The setup is schematically shown in Fig. 3a (see the Methods for details). Figure 3a also shows a typical TA spectrum (expressed in terms of the differential optical density ΔOD divided by the optical density OD). The four graphs in Fig. 3b show the TA dynamics integrated over the energy region near $E_g$ for different QD diameters. A large absorption change can be observed for the larger QDs ($d$ = 3.8 nm and 6.4 nm), while the smaller QDs ($d$ = 2.4 nm and 2.8 nm) exhibit notably smaller changes. Figure 3c shows the average number of carriers per excited volume, $n_d$, estimated from our TA data. The carrier density was determined by comparing the results of the above MIR-pump–white-light-probe TA measurement with those of visible-pump–white-light-probe TA measurements (Supplementary Information IV and V). The QDs with $d \leq 2.8$ nm have very low excited carrier densities compared with those in larger QDs. Hence, the QD size dependence of $n_d$ has the same tendency as that of the HH intensity (Fig. 2). This implies that, with respect to nonlinear responses such as HHG, the primary effect of a larger crystal is more efficient nonlinear carrier generation, and this results in larger HH intensities in the case of solids.

If laser light is applied to a semiconductor, carriers are accelerated by the electric field and their change in behavior in real-space corresponds to an intraband transition in the Brillouin zone or *k*-space. In the case of acceleration of excited carriers in a bulk crystal, they pass through a continuum of states in momentum space. While, for a sample with discrete electronic states (due to quantum confinement), intraband transitions are less likely to occur due to energy gaps between the discrete states. Therefore, we discuss the suppression of HHG from the viewpoint of intraband transitions: To elaborate the impact of quantum confinement on the intraband transitions for HHG, we investigated the optically induced electron dynamics in a simple one-dimensional dimer chain model (Supplementary Information VI). Here, we considered a system Hamiltonian where the different sites of the chain correspond to two species of atoms, such as Cd and Se, and the coupling between the sites is represented by the nearest-neighbour hopping parameter. We decomposed the perturbation part of the Hamiltonian into components corresponding to intra- and interband transitions and calculated the HHG power spectrum $I(\hbar\omega)$ of a single QD.

Figure 4a shows $I(\hbar\omega)$ divided by the QD volume, $I_{\text{HHG}}(\hbar\omega)$, for different QD diameters (chain lengths). As shown in Fig. 4b, we determined the QD size dependence of $I_7$ and $n_d$ by considering different calculation conditions. The red solid and dashed curves are results obtained when both intra- and interband transitions are considered and the QD energy levels depend on *d* (full model). These curves show that $I_7$ and $n_d$ increase abruptly when the QD size becomes larger than 3 nm. On the other hand, when the intraband-transition components in the Hamiltonian are set to zero (Fig. 4b; blue curve, Sec. C of Supplementary Information VI), the intensities are substantilly reduced. This result indicates that the observed nonlinear responses are determined by the size dependence of the intraband transitions. Note that $E_g$ increases as *d* becomes smaller ($\propto 1/d^2$) and the energy gap $\Delta_{\text{sub}}$ between the quantum states with the quantum numbers $n = 1$ and $n = 2$ is roughly proportional to $n^2/m^*d^2$ (Fig. 1a of the Extended Data). To see which parameter governs the size dependence, we also evaluated the size dependence of $I_7$ under the assumption of a size-independent $E_g$. The obtained size dependence (Extended Data Fig. 1b) resembles the red solid curve in Fig. 4b, and this indicates that the actual increase in $E_g$ for smaller QDs does not govern the observed size dependence. To understand the effect of discretization of QDs on the HH intensity, we studied the influence of $m^*$ on the size dependence: Figure 4c shows that the $I_7$ of QDs with small diameters ($d \approx 1$ nm) becomes smaller as the mass becomes smaller. This result is consistent with the experimental results in Fig. 2, since the $m^*$ of CdSe (0.1 $m_0$) is smaller than that of CdS ($m^* = 0.2\ m_0$) (Ref. 23). Moreover, the dependence of $I_7$ on $\Delta_{sub}$, which can be changed by varying $m^*$, is shown in Fig. 1c of the Extended Data. The obtained curves of $I_7$ as a function of $\Delta_{sub}$ are similar regardless of the value of $m^*$. This result shows that the discrete electronic states due to confinement in small QDs suppress intraband transitions.

The good agreement between the experimental QD size dependence and our calculations shows that the intraband transitions cause efficient nonlinear carrier generation and HHG in larger QDs. Nonlinear carrier generation is a coherent optical process and thus involves a superposition of multiple transitions from various VB–CB excitation paths (Sec. D of Supplementary Information VI). It is due to nonlinear coupling between intra- and interband transitions: in addition to the contribution from the pure interband transition terms, the contribution from the coupling terms increases when the intraband transitions are enhanced as schematically described in Fig. 4d and the Extended Data Fig. 2. In the case of coherent excitation of small QDs, the carrier acceleration (intraband transition) is drastically suppressed by discrete electronic states. Meanwhile, larger QDs (bulk form) provide more efficient carrier acceleration and thus cause an additional coupling between intra- and interband transitions, which results in efficient carrier generation and then stronger HHG.

Note that, although our calculations can explain the overall behaviour of the experimental results shown in Figs. 2 and 3, there is a slight quantitative discrepancy between the observed trends in Figs. 2 and 3. To clarify it, we considered the yield ratio, which is the HH intensity per ionization event, $I_7/n_d$. In the Extended Data Fig. 3, we can confirm an increase in the yield ratio of the 7th harmonic as the QD diameter decreases from about 7 to 3 nm. Therefore, with respect to the number of electrons in the CB, the 7th harmonic is generated more efficiently in smaller QDs. In small QDs, which are tightly confined systems, the number of excited carriers is reduced by the amount by which intraband transitions are quenched as the subband-gap energy is increased. On the other hand, the overlap of excited electrons and holes increases in smaller QDs, and thus the probability of recombination increases[21,28]. Therefore, the yield ratio of smaller QDs can be larger.

By using size-controlled QDs, we show that quantum confinement can indeed be used to control intraband transitions and thereby influence the carrier density and HHG. This means that the size of a structure constitutes a parameter that can be used to control extreme nonlinear optical phenomena. Our findings on the control of nonlinear optical phenomena by nanosizing can be used in designing sophisticated petahertz optoelectronic devices that will be implemented at the nanoscale[29,30]. The impact of intraband transitions on carrier exciation shown in our work has important implications for light-driven control of material properties and also for micromachining[31,32], which can be realized, for example, by using two light sources with different wavelengths and a tunable phase offset or different polarization states to manipulate intraband transitions[33].

**Acknowledgements**

Y.K. acknowledges support from the Japan Society for the Promotion of Science (JSPS KAKENHI Grant JP19H05465).

**Author contributions**

K.N. and H.H. carried out the experiments. K.N., H.H., S.A.S., H.T., F.S., G.Y. and Y.K. analysed the data. S.A.S. performed the simulations. M.S., R.S. and T.T. synthesized the quantum dots. H.H. and Y.K. conceived and supervised the project. All authors discussed the results and contributed to the writing of the paper.

**Competing interests:** The authors declare no competing interests.

**Figure Captions**

**FIG. 1 | HHG in CdSe and CdS QD films. a**, Energy level diagram (left), absorption spectra of CdSe QDs with different sizes (middle), and experimental setup including TEM images of CdSe QD films (right). The ODs are normalized by the lowest energy exciton peaks. TEM images of CdSe QD film (3.8 and 14 nm) are shown. **b**, HH spectra of CdSe and CdS QD films with different average QD diameters. The plotted intensities are the measured spectral intensity value divided by the excited volume (i.e., the number of excited QDs times the average volume of a single QD), which was determined by dividing the absorbance at the band edge of the QD film by the absorption cross section per unit volume.

**FIG. 2 | QD size dependence of HHG.** Integrated peak intensity per excited volume $I_h$ as a function of $d$ for different orders. Vertical and horizontal error bars represent the standard deviation of integrated peak intensity and that of diameter. The solid curves are a guide to the eye. We normalized each curve to the value at $d = 3.8$ nm for CdSe QDs ($d=3.5$ nm for CdS QDs). . The excitation power densities were 0.45 and 0.75 TW/cm$^2$ for CdSe and CdS, respectively. The dotted lines indicate the background level. The larger-gap semiconductor CdS has a higher breakdown threshold and can withstand longer exposures. Thus the exposure time for the CdS QDs was four times longer than that for the CdSe QDs (1000 s and 250 s, respectively), which allows the background level (BG) of CdS to be lower than that of CdSe.

**FIG. 3 | TA measurements. a,** Schematic illustration of the experiment and typical TA spectrum of CdSe QDs obtained using an MIR pump pulse (3 μm, 0.36 TW/cm$^2$). **b,** Time evolution of $\Delta$OD/OD near the bandgap energy for $d$ = 2.4, 2.8, 3.8, and 6.4 nm. Here, $\Delta$OD/OD was obtained by measuring the sample transmissivities with and without pump pulse excitation. **c,** Average number of carriers per excited volume, $n_d$, estimated from the

exciton amplitudes in transient absorption signals. Vertical and horizontal error bars represent the standard deviation calculated from the exciton amplitude and that of diameter. The solid curve is a guide to the eye. Since the TA signal of the sample with $d$ = 2.4 nm was smaller than the background level, the point at $d$ = 2.4 nm plots the upper limit of $n_d$ estimated from the background level.

**FIG. 4 | Calculation results. a,** HH spectra per excited volume for different QD diameters (chain lengths). For this calculation, we assumed $\hbar\omega_0$ = 0.35 eV and $E$ = 11 MV/cm, which resembles the experimental conditions. **b,** Diameter dependence of $I_7$ and $n_d$ obtained by using the full model (red circles and squares) and diameter dependence of $I_7$ obtained by using a model excluding the intraband transition term (blue triangles). **c,** Dependence of $I_7$ on diameter for different reduced masses. **d**, The schematic illustration describes nonlinear carrier generation via an additional path including excitation processes due to coupling of intraband and interband transitions.

**Extended Data FIG. 1 |Additional calculation results. a,** Bandgap energy $E_g$ (blue squares) and subband gap, $\Delta_{sub}$ (red circles), as a function of the QD diameter (chain length). **b,** Diameter dependence of $I_7$ obtained by assuming a size-independent $E_g$ (green squares) and that obtained by the full model with a size-dependent $E_g$ (red circles). **c,** Dependence of $I_7$ on $\Delta_{sub}$ for different reduced masses.

**Extended Data FIG. 2 | Schematics of multiple excitation paths.** In addition to the contribution of the pure interband transition terms (left), the efficient intraband transition in larger QDs (or bulk) opens multiple excitation paths due to the nonlinear coupling between the intra- and interband transitions (right). These additional excitation channels due to the coupling promote nonlinear carrier injection and enhance HHG in larger QDs.

**Extended Data FIG. 3 | Yield ratio.** Diameter dependence of the yield ratio of the 7th order for CdSe, $I_7/n_\mathrm{d}$. The data is normalized to the value at $d = 6.4$ nm.

**Methods**

**QD synthesis.** The CdSe and CdS QDs were prepared in our laboratory by wet chemical synthesis, which provides precise control of the QD size (see Supplementary Information I for the details).

**HHG spectrum measurements.** For the measurement of the HH spectra, QD thin films were excited by linearly polarized MIR pulses with $\hbar\omega_0 = 0.35$ eV ($\lambda_0 = 3.5$ μm). The MIR pulses were generated using a multi-stage optical parametric amplifier (OPA) system (OPerA with the NDFG1 option, Coherent) driven by a Ti:sapphire laser with a repetition rate of 1 kHz. The QD thin film samples were prepared by spin-coating of colloidal QDs dispersed in hexane on a sapphire substrate with a thickness of 0.5 mm. For all QD samples, hexane was evaporated after the spin-coating on the substrate, and it was confirmed that the substrate and the surface-protecting ligands for QD do not contribute to the HH spectra. The experiments were conducted with excitation intensities below the damage threshold of the QDs; the excitation peak intensities were 0.45 TW/cm$^2$ for the CdSe QDs and 0.75 TW/cm$^2$ for the CdS QDs (the pulse width was 80 fs and the spot diameter was ≈ 300 μm.) We verified that the linear absorption at the excitation spot did not change during the MIR-pulse irradiation and that the observed change in $I_{HHG}$ for smaller QDs was not due to sample damage (Fig. S3 of Supplementary Information). All experiments were performed at room temperature. The HHs were detected using a charge-coupled-device (CCD) camera (PIXIS, Teledyne Princeton Instruments) attached to a spectrometer (SpectraPro, Teledyne Princeton Instruments).

**Transient absorption spectroscopy.** In the MIR-pump–visible-probe TA measurements, we used an MIR excitation pulse with $\hbar\omega_0 = 0.41$ eV ($\lambda_0 = 3$ μm, the temporal pulse width was 100 fs), which is generated from a home-made OPA system based on an Yb: KGW femtosecond laser system (Pharos, Light Conversion) with a center wavelength of 1033 nm, pulse width of 180 fs, and 2 mJ/pulse at a repetition frequency of 1 kHz. To avoid the additional small thermal-decay contribution in the TA signal due to the ligands of the QDs, it was necessary to use an MIR excitation wavelength (3 μm) that is shorter than the one used in the HH spectrum measurements shown in Fig. 1. We verified that the different MIR excitation wavelength did not cause a notable change in the size dependence of $I_h$, as shown in Fig. S4 of the Supplementary Information. For the probe pulses, white light was generated by focusing a small fraction of the output beam of the Yb: KGW laser into a 10-mm-thick quartz cell containing water (see Supplementary Information V for the details). This method provided a better signal-to-noise ratio compared to using the Ti:sapphire laser owing to better laser stability. The generated continuum spectrum ranged from the visible to the near infrared. The MIR pump and visible probe beam were focused onto the sample and the beam spot diameter was about 250 μm and 150 μm. For the TA measurements with visible pump pulses [$\hbar\omega_{vis}$ = 2.40 eV ($\lambda_0$ = 517 nm) and the temporal pulse width was 180 fs], we used frequency-doubling of the fundamental pulses in a beta barium borate (β-BBO) crystal. Here, we used QDs with $d$ = 2.4, 2.8, 3.1, 3.8, 4.5, and 6.4 nm, because the band-edge exciton energies were respectively 2.40, 2.31, 2.24, 2.19, and 1.91 eV, allows a band-to-band excitation by the second harmonic of the fundamental pulse ($\hbar\omega_{vis}$ = 2.40 eV). The probe light transmitted through the sample was spectrally resolved by a spectrometer and detected using a CCD camera. To obtain a high signal-to-noise ratio, we synchronously chopped the pump beam at 500 Hz, thereby blocking every second pump pulse. A computer collected the signals to obtain the TA signal, which was calculated from the probe transmission with and without the influence of the pump light.

**Data availability**

Source data are available for this paper. All other data that support the plots within this paper and other findings of this study are available from the corresponding author upon reasonable request.

**Figure 1**

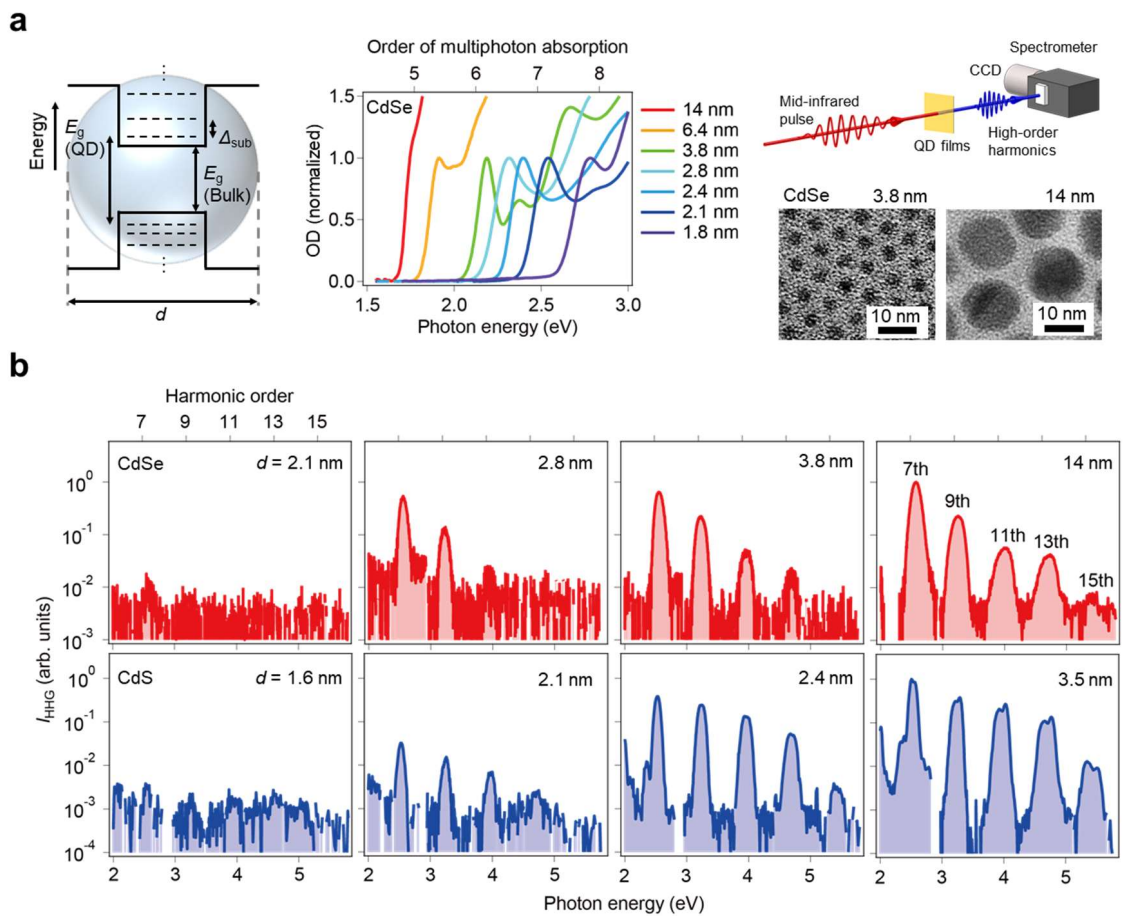

K. Nakagawa

**Figure 2**

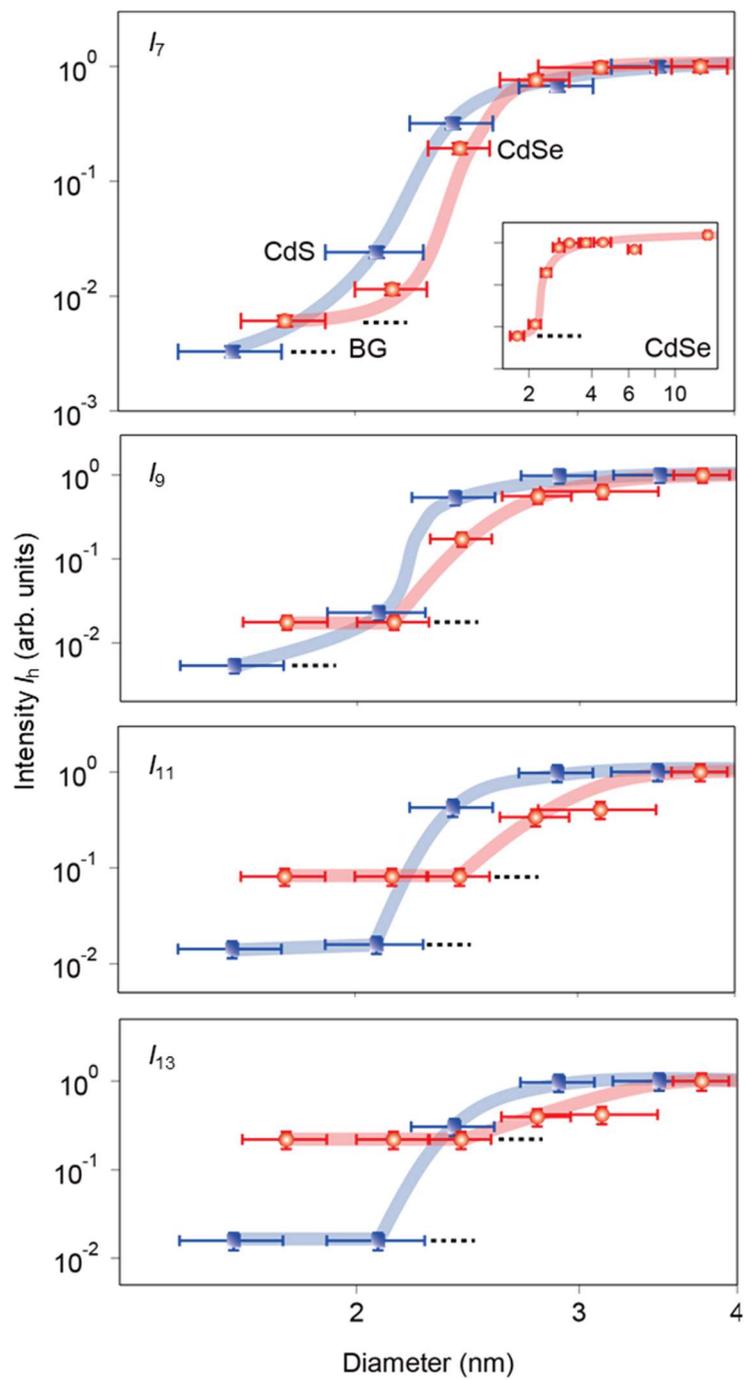

**K. Nakagawa**

**Figure 3**

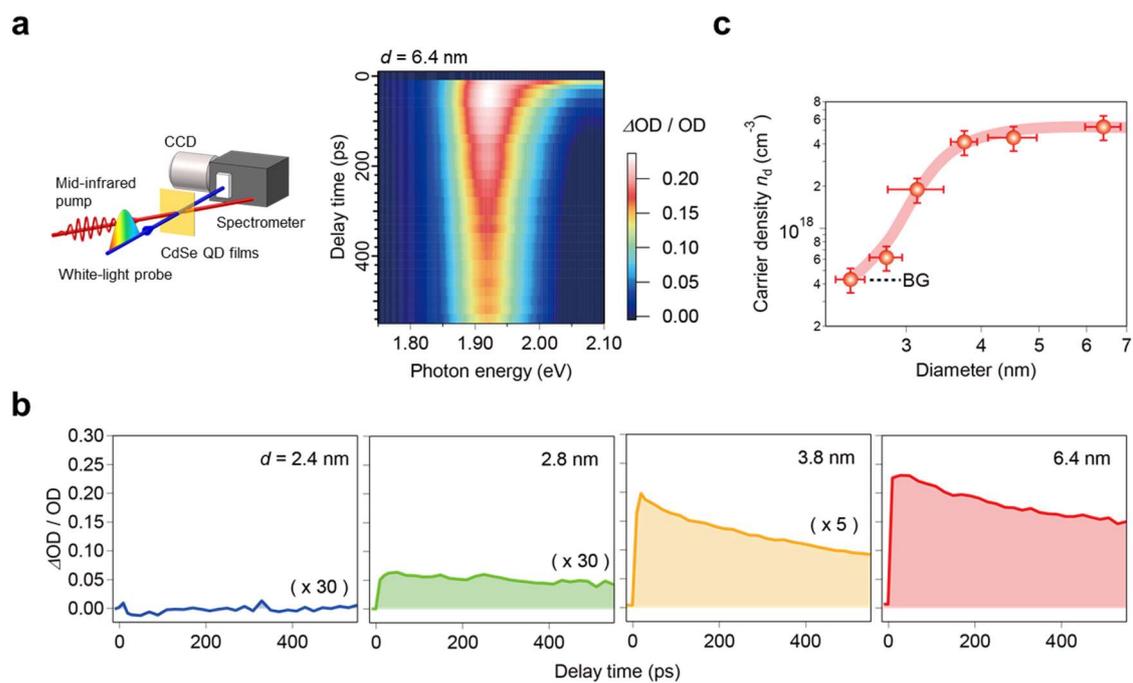

K. Nakagawa

**Figure 4**

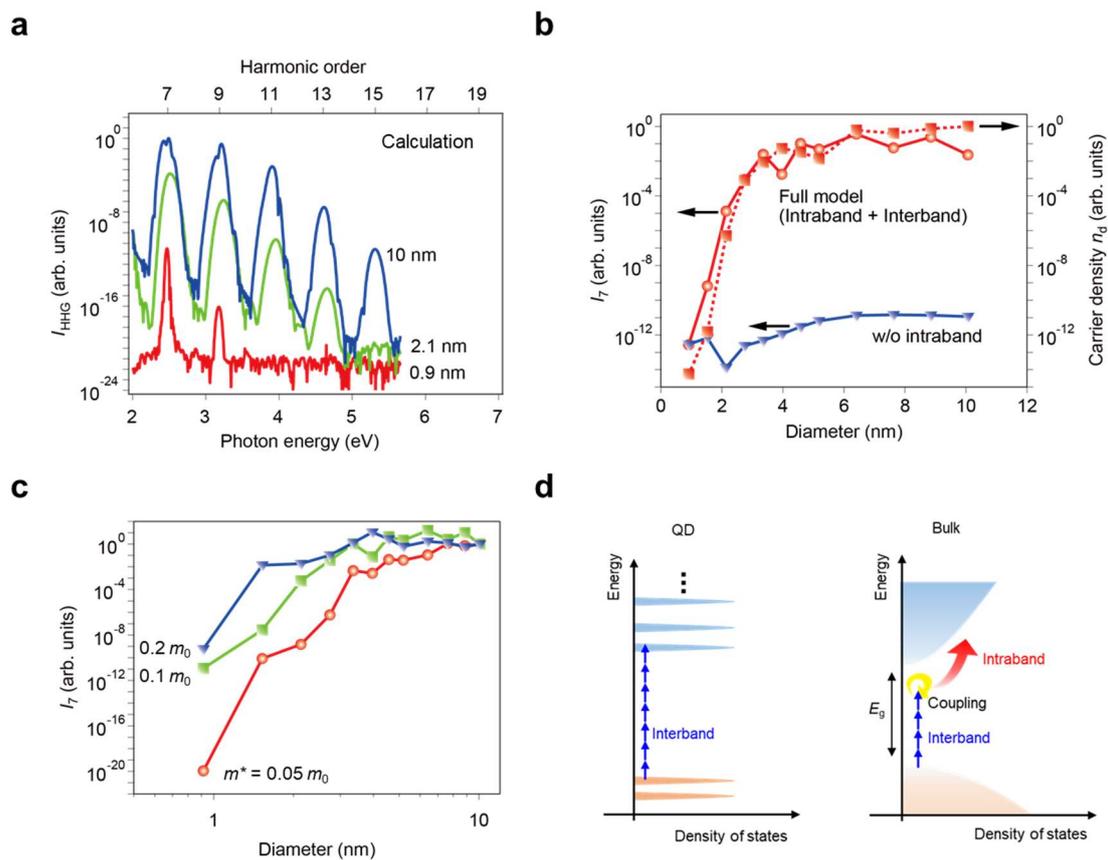

**K. Nakagawa**

**Extended Data Figure 1**

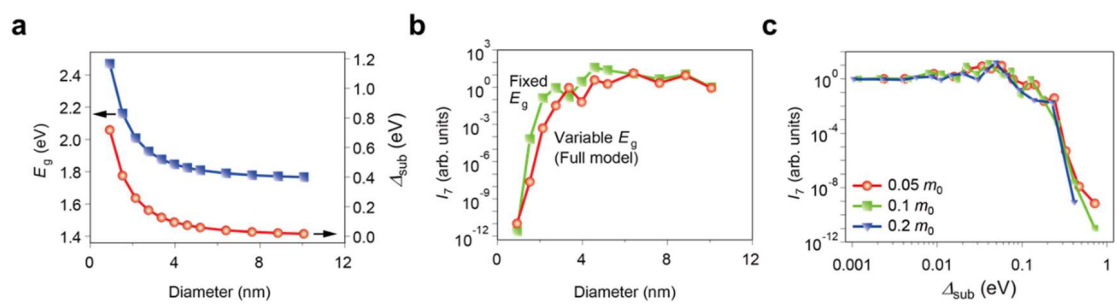

K. Nakagawa
23

**Extended Data Figure 2**

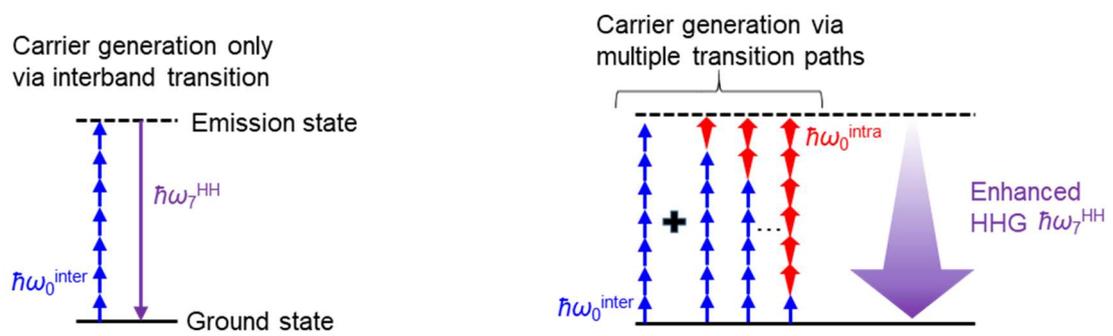

K. Nakagawa

**Extended Data Figure 3**

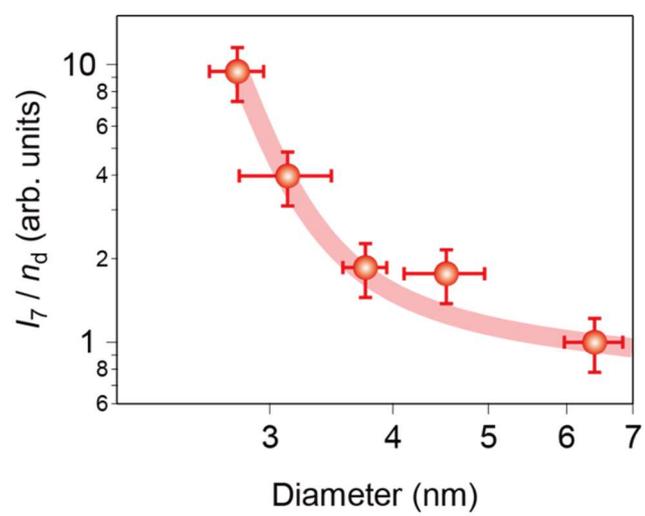

**K. Nakagawa**